\documentclass[11pt,twoside]{article}

\usepackage{asp2006}
\usepackage{epsf}
\usepackage{psfig}
\usepackage{lscape}

\usepackage{adassconf}

\usepackage{graphicx}
\usepackage{sidecap}

\begin{document}
\title{Structure and Evolution of the Opacity of Spiral Disks}   
\author{B. W. Holwerda\altaffilmark{1}, R. A. Gonz\'{a}lez\altaffilmark{2}, W. C. Keel\altaffilmark{3}, D. Calzetti\altaffilmark{4}, R. J. Allen\altaffilmark{1} and P. C. van de Kruit\altaffilmark{5}}   

\altaffiltext{1}{Space Telescope Science Institute} 
\altaffiltext{2}{UNAM}
\altaffiltext{3}{University of Alabama}
\altaffiltext{4}{University of Massachusetts}
\altaffiltext{5}{Kapteyn Astronomical Institute }

\begin{abstract} 
The opacity of a spiral disk due to dust absorption influences every measurement we make of it in the UV and optical. Two separate techniques directly measure the total absorption by dust in the disk: calibrated distant galaxy counts and overlapping galaxy pairs. The main results from both so far are a semi-transparent disk with more opaque arms, and a relation between surface brightness and disk opacity. In the Spitzer era, SED models of spiral disks add a new perspective on the role of dust in spiral disks. Combined with the overall opacity from galaxy counts, they yield a typical optical depth of the dusty ISM clouds: 0.4 that implies a size of $\sim$ 60 pc. Work on galaxy counts is currently ongoing on the ACS fields of M51, M101 and M81. Occulting galaxies offer the possibility of probing the history of disk opacity from higher redshift pairs. Evolution in disk opacity could influence distance measurements (SN1a, Tully-Fisher relation). Here, we present first results from spectroscopically selected occulting pairs in the SDSS. The redshift range for this sample is limited, but does offer a first insight into disk opacity evolution as well as a reference for higher redshift measurements.
\end{abstract}



The opacity of spiral disks is a characteristic that influences many of our observations of disks at any redshift. Thus far, two observational methods have produced reliable measurements of disk opacity; occulting galaxy pairs and the calibrated number of more distant galaxies. Together with {\em Spitzer} SED models, the apparent optical depth can be used to estimate general properties of the extincting ISM 's structure.


\begin{SCfigure}
  \centering
  \includegraphics[width=0.5\textwidth]{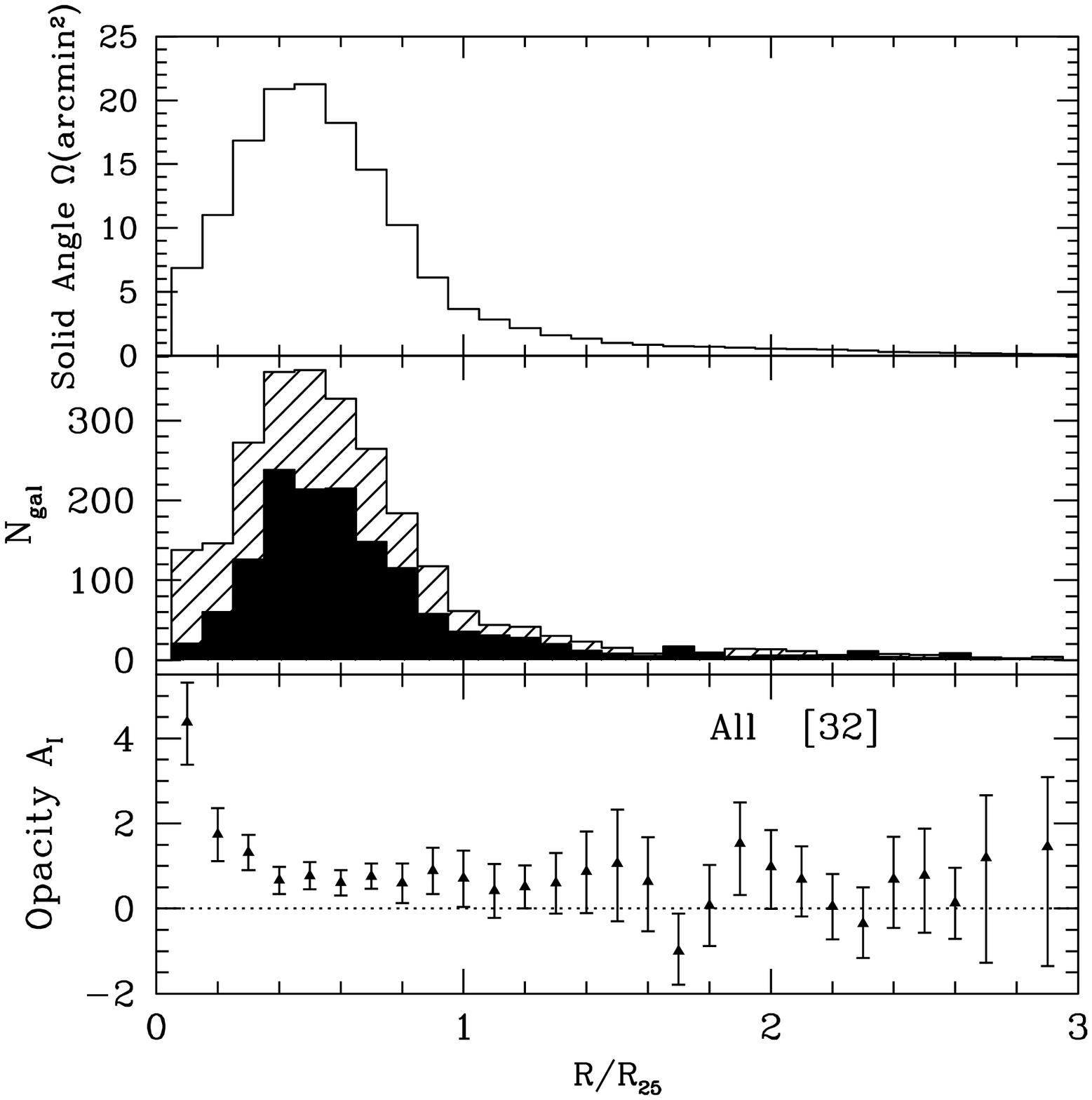}
  \caption{Radial opacity profile from \protect\cite[][bottom panel]{Holwerda05b}, the numbers of real and artificial galaxies on which the opacity measurement is based (middle), and the combined solid angle in each radial bin (top). Galaxy counts at higher radii (for instance, the GHOSTS data, de Jong et al, this volume), and large mosaics on single disks will allow us to better characterize the point where disks become completely transparent.}
  \label{f:ra}
\end{SCfigure}

\section{Distant Galaxy Counts}

The number of distant galaxies seen in an {\em HST} image through the face-on foreground spiral is a direct indication of its opacity, after proper calibration using artificial galaxy counts \citep[synthetic field method (SFM)][]{Gonzalez98,Holwerda05a}. We have obtained calibrated galaxy counts for a sample of 32 deep {\em HST/WFPC2} fields. The main results from the disk opacity study are: (1) most of the disks are semitransparent \citep[][Figure \ref{f:ra}]{Holwerda05b}, (2) arms are more opaque \citep{Holwerda05b}, (3) as are brighter sections of the disk \citep{Holwerda05d}. We did not find a relation between HI profiles and disk opacity but this can be better explored on a single disk \citep{Holwerda05c}. The optimal distance of the foreground disk is $\sim$ 10 Mpc, the compromise between crowding effects and solid angle constraints \citep{Gonzalez03,Holwerda05e}.

\begin{SCfigure}
  \centering
  \includegraphics[width=0.5\textwidth]{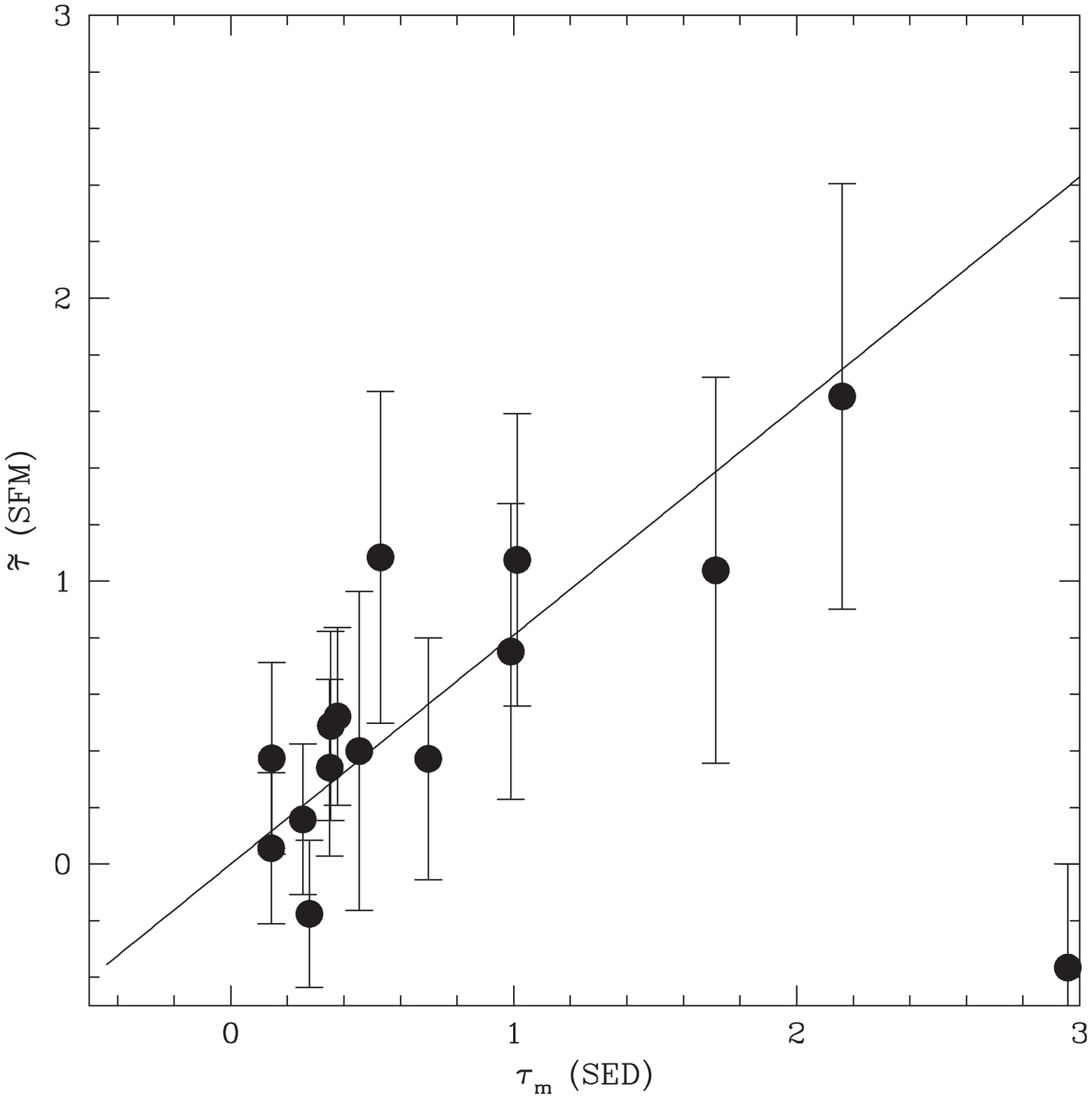}
  \caption{The relation between the optical depth from the SED model [$\tau_m$(SED)], and the observed optical depth from the number of distant galaxies [$\tau$(SFM)]. Model C from \protect\cite{Natta84} is fit through these. Cloud optical depth is 0.4, and  more than a single cloud along the line-of-sight is needed, especially for optically thick disks. Figure from \protect\cite{Holwerda07a}.}
  \label{f:tc}
\end{SCfigure}


There is an overlap of 12 galaxies with {\em SINGS}; we compared the disk opacity from number of distant galaxies to the dust surface density obtained from an SED model \citep{Draine07a}.
Expressed as optical depths, the relation between the {\it observed} optical depth ($\tau$) and the optical depth inferred from the SED dust mass ($\tau_m$) is solely a function of the typical cloud optical depth ($\tau_c$): $\tau/\tau_m = (1-e^{-\tau_c})/\tau_c$ \citep{Natta84}. Figure \ref{f:tc} shows the values for $\tau$ and $\tau_m$ and the fit of $\tau_c$=0.4. This implies that most of the disks are made up of small, optically thin, cold ISM structures with more than one cloud along the line-of-sight, especially in optically thick disks \citep{Holwerda07a}. From their distribution in the E[I-L] color map, it becomes clear that the number of distant galaxies does not drop exclusively as a result of grand spiral opaque structures but that the unresolved dusty ISM disk is equally important \citep{Holwerda07b}.


\begin{figure}
  \centering
  \includegraphics[width=0.6\textwidth]{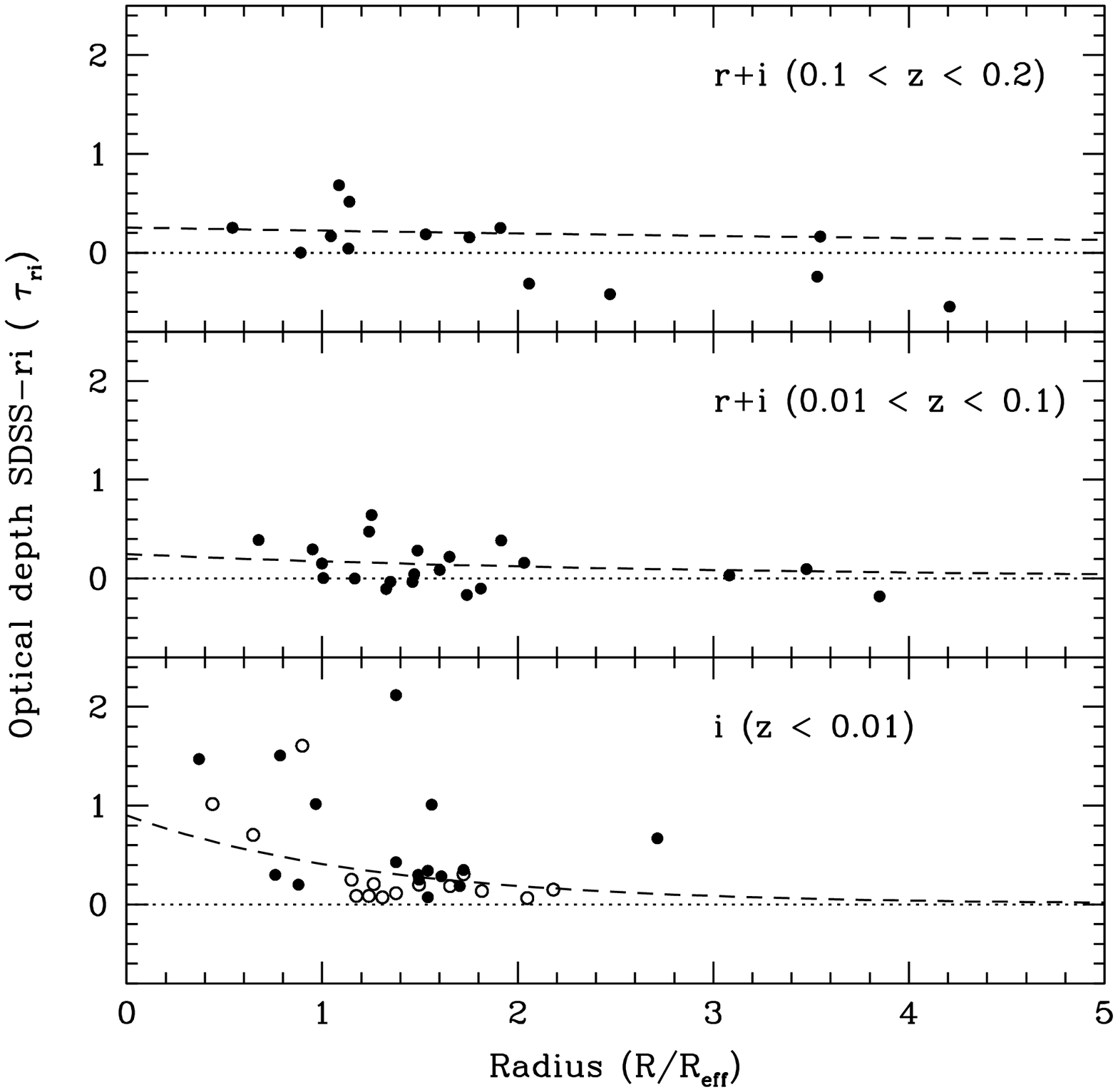}
\caption{\label{f:raz}Radial opacity profile for the local (lower), z=0-0.1 (middle) and z=0.1-0.2 (top) spirals from occulting pairs. Open circles are disk sections for the local pairs. Figure from \protect\cite{Holwerda07c}.}
  \label{f:tt}
\end{figure}

\section{Occulting Galaxies}

A galaxy partially occulted by a face-on spiral galaxy can be used to determine the opacity of the foreground spiral. The method has seen extensive use on nearby pairs, using ground-based imaging 
\citep{Andredakis92, Berlind97, kw99a, kw00a}, spectroscopy \citep{kw00b} and, more recently, HST imaging \citep{kw01a, kw01b, Elmegreen01}. Results from these included radial plots of extinction all the way through the disk, showing gray extinction when taken over large sections, but with a 
near-galactic reddening law for scales smaller than 100 pc, when resolved with {\em HST}.

The Sloan Digital Sky survey is a powerful tool for finding rare objects such as the ideal occulting pair, a face-on spiral partially covering a more distant elliptical. From the SDSS spectra, we selected 86 galaxy pairs, with emission spectra at lower redshift on top of an elliptical spectrum \citep[similar to][]{Bolton04}.
We have applied the occulting galaxy method on the SDSS images, obtaining an opacity measurement at a single radius for many of these pairs. 
Figure \ref{f:raz} shows the radial opacity plot for local spirals, and for two more distant redshift bins from the SDSS. Close to the center of the spirals there is not enough flux from the background elliptical, but the radial profiles at z = 0-0.1 and z = 0.1-0.2 are similar to the local one for a mix of arm and disk values. 

{\em HST} imaging of these pairs would allow us to (a) measure the opacity closer to the spiral's centre and (b) distinguish between arm and disk regions, and trace the radial opacity profile fully.

\section{Future Outlook}

One can now confidently state that we know the dust content and the resulting extinction in local spiral disks within the optical radius. However, two issues remain: how far outside the optical disk can we detect the effect of dust and how much has the opacity of spiral disks changed since z $\sim$ 1, when star formation was an order of magnitude higher? The radial extent can be explored with galaxy counts in large {\em HST} mosaics of local disks or with deep FIR {\it Spitzer} observations (see Hintz, this volume), and the evolution of disk opacity with large numbers of occulting pairs from both SDSS (Keel et al., this volume) and {\em HST} deep surveys.





\end{document}